# Interaction between skyrmions and antiskyrmions in a coexisting phase of a Heusler material

Daigo Shimizu[1]*, Tomoki Nagase[1], Yeong-Gi So[2], Makoto Kuwahara[3], Nobuyuki Ikarashi[1,4], and Masahiro Nagao[1,4]*

[1] Department of Electronics, Graduate School of Engineering, Nagoya University, Nagoya, Japan.
[2] Department of Materials Science, Graduate School of Engineering Science, Akita University, Akita, Japan.
[3] Advanced Measurement Technology Center, Institute of Materials and Systems for Sustainability, Nagoya University, Nagoya, Japan.
[4] Center for Integrated Research of Future Electronics, Institute of Materials and Systems for Sustainability, Nagoya University, Nagoya, Japan.
e-mail: shimizu.daigo@d.mbox.nagoya-u.ac.jp; nagao.masahiro@imass.nagoya-u.ac.jp

**Coexisting phases of magnetic skyrmions and antiskyrmions have proposed to exhibit a variety of fascinating properties, owing to interactions between them. The recent discovery of the coexisting phase in a Heusler material could offer a platform for skyrmion-antiskyrmion-based spintronics. Here we report Lorentz electron microscopy experiments and micromagnetic simulations in a similar Heusler material, $Mn_{1.3}Pt_{1.0}Pd_{0.1}Sn$. Around $B_c$~420 mT, we find a stochastic reversible transformation and a room temperature coexisting phase of elliptical skyrmions and square-shaped antiskyrmions. The closeness of the energy competition is sensitive to the exchange stiffness constants and sample thickness. Furthermore, we reveal isotropic long-range repulsive interaction between the skyrmions and antiskyrmions regardless of their shapes and the skyrmion helicities, in stark contrast to conventional thought of angle- and helicity-dependent short-range pairwise interactions. The observed interaction possibly results from the topological protection against the intrusion of magnetic flux density coming from skyrmions (antiskyrmions) into antiskyrmions (skyrmions). Our results provide new insight into interacting skyrmions and antiskyrmions and a guide for developing skyrmion-antiskyrmion-based spintronics.**

Skyrmions have been ubiquitous in condensed matter, whereas antiskyrmions rarely exist alone. Theoretical studies have discussed the properties involved in the short-range pairwise interactions between skyrmions and antiskyrmions in Bose-Einstein condensates[1], quantum Hall systems[2] including bilayer graphene[3,4,5]. In magnets, experiments have directly observed skyrmions and antiskyrmions as single phases, respectively[6,7], that offer appealing platforms to explore emergent electromagnetic fields and related properties. Magnetic skyrmions and antiskyrmions are non-coplanar spin structures with opposite topological charge[8]. Skyrmions are stabilized by Dzyaloshinskii-Moriya interaction (DMI) originating from spatial inversion symmetry breaking of the systems such as chiral magnets[6] and multilayer ultrathin films[9]. On the other hand, antiskyrmions are only realized in Heusler materials[7] with $D_{2d}$ symmetry (Fig. 1a) leading to anisotropic DMI (Fig. 1b).

Of particular interest are coexisting phases of magnetic skyrmions and antiskyrmions, where recent many theories have proposed a variety of fascinating properties[8,10-16] that do not appear in each single phase. Examples include skyrmion-antiskyrmion liquid[10], rectangular lattice with alternating columns of skyrmions and antiskyrmions[11], phase separation[12,13], current-induced pair annihilation that emits propagating spin wave[14], topological conversion by their collision[8,14,15], and rectilinear and trochoidal motions of skyrmions and antiskyrmions, respectively[16]. Recent theories have predicted the coexisting phases in centrosymmetric frustrated systems[10-15]. However, no experimental evidence has been presented[17]. Meanwhile, systems with anisotropic DMI have been in conflicting discussions that are whether elliptical skyrmions or antiskyrmions are stable[18,19,20]. In the heat of the theoretical contradictions, recent Lorentz transmission electron microscopy (L-TEM) experiment has discovered the coexisting phase[21] at 268 K in the inverse tetragonal Heusler material, $Mn_{1.4}Pt_{0.9}Pd_{0.1}Sn$. At the same time, in $Mn_{1.4}Pt_{0.9}Pd_{0.1}Sn$ thin film, another L-TEM experiment has demonstrated a one-way topological transformation from square-shaped antiskyrmions to elliptical Bloch skyrmions with clockwise (CW) and anticlockwise (ACW) via the non-topological (NT) bubbles by in-plane magnetic fields, as the following procedure[22]. When the incident electron beam (parallel to the magnetic field) is parallel to the [001] direction, the tilt angle is $\theta = \sim 0°$. The rotation axes are the [100] directions. When the sample tilts away from $\theta = \sim 0°$ where square-shaped antiskyrmions are observed, an in-plane component of the magnetic field induces bullet-shaped NT-bubbles with two Bloch lines, and then back to $\theta = \sim 0°$, NT-bubbles convert to elliptical Bloch skyrmions with clockwise (CW) or anticlockwise (ACW) depending on the rotation axes. These discoveries could make Heusler materials a platform for skyrmion-antiskyrmion-based spintronics. Although the interaction between the skyrmions and antiskyrmions governs current-induced dynamic properties, its nature remains unrevealed.

Here we report L-TEM experiments and micromagnetic simulations in similar Heusler material, $Mn_{1.3}Pt_{1.0}Pd_{0.1}Sn$. This composition is different from $Mn_{1.4}Pt_{0.9}Pd_{0.1}Sn$ as previously reported[7,21-27]. The reason is that we initially intended to synthesize $Mn_{1.4}Pt_{0.9}Pd_{0.1}Sn$, but the composition has somehow shifted slightly in our synthesis. At the magnetic fields of $B_c = 420 \pm 40$ mT and a thickness of $t \sim 100$ nm, we show a stochastic reversible transformation between elliptical skyrmions and square-shaped antiskyrmions via NT-bubbles with two Bloch lines. This transformation occurs by an in-plane component of the magnetic field opposite to two Bloch lines of NT-bubbles. Besides, we find a room temperature (RT) coexisting phase. Our micromagnetic simulations suggest that the closeness of the energy competition between skyrmions and antiskyrmions is sensitive to the exchange stiffness constants and sample thickness. Furthermore, we observe signatures of isotropic long-range repulsive interactions between skyrmions, antiskyrmions, and themselves, by exploiting annihilation of the spin textures that have been performed by local heating with a focused electron beam. Micromagnetic simulations support the experimental interpretation and suggest that isotropic long-range repulsive interactions between skyrmions and antiskyrmions result from preventing topological destruction against the intrusion of leak magnetic flux density into skyrmions and antiskyrmions. In addition, antiskyrmions gradually turn from square lattice to disordered state with increasing the magnetic field ($B$), the latter appearing at ~450 mT which is around a little below the upper limit of $B_c$. This behaviour indicates that antiskyrmion-antiskyrmion interaction turns from anisotropic to isotropic.

We find the stochastic reversible transformation between antiskyrmions and skyrmions via NT-bubbles and the coexistence at $B_c$ = 420 ±40 mT. All of the following L-TEM experiments were performed at RT and the direction of $B$ along to the incident electron beam, the sample tilt angle is $\theta$, and the rotation axes are in the [100] directions (Fig.1c). Figure 1d-h shows a series of L-TEM images in $Mn_{1.3}Pt_{1.0}Pd_{0.1}Sn$ thin film with $t \sim 100$ nm. A square-shaped antiskyrmion appears as indicated by the yellow arrow in Fig. 1d. We have produced antiskyrmions by tilting the sample to $\theta$ = 15° and then to $\theta = -0.7°$ via $\theta = -2.1°$ which is the same procedure as in the previous study[21]. There are some NT-bubbles, and the reasons for this are discussed below. In TEM imaging principle, when an incident electron beam is precisely parallel to the crystal axis, the TEM image becomes pitch-black due to diffraction contrasts which obscure the magnetic ones (Supplementary Fig. 1), indicating that L-TEM imaging is not able to observe magnetic contrasts at $\theta = 0°$. By tilting the sample at $\theta = +15°$, the antiskyrmion transforms into NT-bubble (Fig. 1e). We subsequently tilted the sample from $\theta = +15°$ to $\theta = -2.1° \sim -2.3°$ passing $\theta = 0°$, and then to $\theta = -0.6° \sim -0.8°$. The latter procedure was done to avoid diffraction contrasts, the so-called bend contours where the Bragg condition is locally satisfied due to the sample bend, that partially cover local regions to be observed. We observe the ACW elliptical skyrmion (Fig. 1f). This one-way transformation in itself is consistent with that in the previous study[22]. By performing the same procedure again, the antiskyrmion reappears via NT-bubble (Fig. 1g,h). These results indicate that this kind of Heusler materials allows not only one-way[22] but also reversible transformation. It is noted that the transformation stochastically occurs. Similar stochastic reversible transformations are observed for CW elliptical skyrmions by tilting the sample around the [010] axis.

Although the previous study[22] has discussed the energetic stability of antiskyrmions and skyrmions, the transformation mechanism has not yet been clarified. Here we show the mechanism of stochastic reversible transformations, resulting from stochastically creating two Bloch lines into NT-bubble or vanishing two Bloch lines of NT-bubble by negative in-plane magnetic fields. Magnetic symmetry breaking and restoration occur during stochastic reversible transformations. Figure 1i displays three-dimensional magnetic point groups that the spin structures have in order of higher symmetry from the top. Antiskyrmions, skyrmions, and NT-bubbles have magnetic point groups of –4'2'm, 2'2'2, and 2', respectively. NT-bubbles, the intermediate states, have the lowest symmetry. Symmetry restoration occurs at the transformation from NT-bubbles to skyrmions (Fig. 1e-f) by restoring a symmetry element of 2', and from NT-bubbles to antiskyrmions (Fig. 1g-h) by restoring that of –4', 2', and m. Negative in-plane magnetic fields opposite to the net in-plane magnetic component of two Bloch lines of NT-bubble induce symmetry restorations from NT-bubble. Figure 1j shows a schematic of the transformation from NT-bubble (the left) with two Bloch lines (the small blue arrows) to other spin structures (the right column) by a negative in-plane magnetic field (the large red arrow denoted by $B_{in}$). In the experiment, slightly tilting the sample to $\theta < 0°$ (Fig. 1f,h) induces the negative in-plane magnetic fields. When the negative in-plane magnetic field stochastically creates additional two Bloch lines (the small red arrows) of which the net in-plane component is parallel to the negative in-plane magnetic field, the antiskyrmion with four Bloch lines in NT-bubble is formed (the yellow arrow in Fig. 1h, top of the right column in Fig. 1j). On the other hand, when stochastically vanishing two Bloch lines, of which the net in-plane component is antiparallel to the negative in-plane magnetic field, of original NT-bubble, the skyrmion without any Bloch lines in NT-bubble is formed (the yellow arrow in Fig. 1f, second of the right column in Fig. 1j). When the creation and annihilation of Bloch lines simultaneously occur, they change into another NT-bubble (the black arrows in Fig. 1f, third of the right column in Fig. 1j). In the other case, NT-bubble remains unchanged (the white arrow in Fig. 1h, bottom of the right column in Fig. 1j). In applications, a local in-plane pulsed magnetic field would allow deterministic switching between skyrmions and antiskyrmions.

Our micromagnetic simulations show that skyrmions, antiskyrmions, and NT-bubbles have topological charges of $N = +0.992$, $–0.972$, and $+0.007$, in the presence of an in-plane magnetic field corresponding to $\theta = –3°$, and of $N = +0.993$, $–0.974$, and $+0.009$ in the absence of an in-plane magnetic field. The differences between the former and the latter are so slight. We can conclude that the topology has not essentially changed. These values deviate slightly from the ideal values ($N = \pm 1, 0$), because of the remaining components of the helical spin structure at low magnetic fields, and they approach the ideal values as the magnetic field increases.

We also observe the RT coexisting phase (Fig. 2a), by once tilting the sample to $\theta = +15°$ and then to $\theta = –1.2°$ via $\theta = –2.1°$. The core-to-core distance between skyrmions and antiskyrmions is $d_{Sk\text{-}aSk} = \sim 335$ nm. The temperature enhancement of the coexisting phase may be due to the composition shift. From the viewpoint of the crystal structure with $D_{2d}$ symmetry, as Mn elements are in excess or deficient, either Mn site of the two Mn sublattices introduces antisite defects or chemical disorders. As a result, Heusler materials including Mn-Pt-Pd-Sn compounds show characteristics that make it easy to significantly change the magnetic properties with small changes in the magnetic elements[28]. Although it is difficult to clarify the chemical composition dependent magnetic properties of Mn-Pt-Pd-Sn in the present study, these suggests that the chemical composition is responsible for the presence of the RT coexisting phase in $Mn_{1.3}Pt_{1.0}Pd_{0.1}Sn$. In the upper region, many NT-bubbles exist, and the direction of their bullet shape remains unchanged from that at $\theta = +15°$. In this region, bend contours (strong dark contrasts) exist due to the sample bend, indicating that the in-plane magnetic field is insufficient for changing from NT-bubbles to the other spin textures. Although the L-TEM experiments appear to be strongly competing state of skyrmions and antiskyrmions, it is not clear whether Heusler materials are able to have the closeness of the energy competition in the coexisting phase because of going through the intermediate states (NT-bubbles) in the experiments[7,21-23,27]. Besides, the formation conditions of the coexisting phase have not yet been clarified. Using micromagnetic simulations, we show that the degree of the energy competition depends on the exchange stiffness constant of $A$ including exchange interaction of $J$ and the sample thickness of $t$. First, we used a thin film with $t = 100$ nm and the magnetic parameters of Ref. 22 including $A = 8.0 \times 10^{-12}$ J/m, which can reproduce other experimental results that we observed (see Method and Supplementary Fig. 2, 3, 4.). Figure 2b presents energy density as a function of the normalized magnetic field ($B/B_0$) along the [001] direction, which subtracts ferromagnetic (FM) energy density from skyrmion and antiskyrmion energy densities, denoted by $\varepsilon_{sk}$ and $\varepsilon_{ask}$, respectively. The upper limit of the stable helical state is $B/B_0 = 13.239$ (Supplementary Fig. 3). At $B_{c1}/B_0 = 13.283$, $\varepsilon_{sk}$ and $\varepsilon_{ask}$ have the same energy density. Above $B/B_0 = 13.461$, field-induced FM state is the most stable. In the inset, the difference between $\varepsilon_{sk}$ and $\varepsilon_{ask}$ is less than 0.33 J/m$^3$ around $B_{c1}/B_0 \pm 0.05$, indicating that skyrmions and antiskyrmions are energetically competing.

We found that as the exchange stiffness constant changes, the points of $B_{c1}/B_0$ where $\varepsilon_{sk}$ and $\varepsilon_{ask}$ have the same energy density approach regions of the magnetic field where helical and ferromagnetic states are stable. Comparing the previous experimental studies[22,24-26], the nominal composition of $Mn_{1.4}Pt_{0.9}Pd_{0.1}Sn$ has different Curie temperature of $T_C$ in the range of 375 K to 400 K. This indicates that the change of $T_C$ is sensitive to compositional differences leading to antisite defects and chemical disorders even within the error range of composition measurement in this kind of Heusler materials. In general, $T_C$ of high-temperature materials is proportional to $A$ ($J$). We investigate the effect of varying the exchange stiffness constant on the energy competition. We fixed a film thickness of $t = 100$ nm as well as Fig. 2b and used the exchange stiffness constant of ±10% change from $A = 8.0 \times 10^{-12}$ J/m of Fig. 2b as $A = 7.2 \times 10^{-12}$ J/m and $8.8 \times 10^{-12}$ J/m. At a lower exchange stiffness constant of $A = 7.2 \times 10^{-12}$ J/m, $B_{c1}/B_0$ shift to a lower magnetic field of stable helical phase (Fig. 2c). Contrarily, at a higher exchange stiffness constant of $A = 8.8 \times 10^{-12}$ J/m, $B_{c1}/B_0$ shift to a higher magnetic field of $B_{c1}/B_0 = 14.455$ in the vicinity of stable ferromagnetic phase (Fig. 2d).

Our simulations demonstrate that the sample thickness also changes the energy stability. The previous study[27] reported that the antiskyrmion size and their lattice constants changes as a function of the sample thickness of $Mn_{1.4}Pt_{0.9}Pd_{0.1}Sn$ and $Mn_{1.4}PtSn$. Although the energy stability of NT-bubbles is insensitive to the sample thickness[23], that of skyrmions and antiskyrmions could be sensitive in this kind of Heusler materials. Figure 2e shows the L-TEM image obtained by a region of $t = \sim 130$ nm. Most of the spin structures are antiskyrmions. Micromagnetic simulations support the thickness tendency that the sample thickness changes the energy stability of skyrmions and antiskyrmions. Figure 2f presents the energy density in a thin film of $t = 150$ nm. For any value of the magnetic field and exchange stiffness constant, the antiskyrmion is always energetically lower than the skyrmion. (Micromagnetic simulations in $t = 75$ nm also show the same result as that in $t = 150$ nm, as shown in supplementary Fig. 5.) Our results suggest that Heusler materials are able to have the closeness of the energy competition in the coexisting phase, and in our experiment (Fig. 2a), both the sample thickness and the exchange stiffness constant (chemical composition) have fortuitously appropriated. Adjusting these parameters could allow to realize coexisting phases in other Heusler materials. In the future, detailed further studies are necessary to elucidate why the coexisting phase appears at RT by investigating composition dependence of magnetic properties in Mn-Pt-Pd-Sn and the combination of developing the magnetic phase diagram and simulating the thermodynamic model.

Fortuitously observing RT coexisting phase motivates us to investigate the interactions governing the dynamic properties. In particular, it would be interesting in various physics fields to see whether short-range pairwise interactions between skyrmions and antiskyrmions are universal or not. Here we exploit the local heating with a focused electron beam (Fig. 3a) for annihilating the spin textures according to the theory[29] and then pursue behaviours of the remaining spin textures driven by the interactions. Before the experiments for the coexisting phase, we preliminary demonstrate a local annihilation and examine the thermal current (magnon current) effect on the spin textures. Figure 3b,c shows the antiskyrmion annihilation. As we expose the antiskyrmion to the focused electron beam with a diameter of $r \sim 200$ nm in 2 minutes (Fig. 3b), the antiskyrmion surely vanished (Fig. 3c). Local heating generates magnon current simultaneously with thermal current because a radial temperature gradient occurs from a local heating position in our experiment. Theories predicted that magnon current provides a spin transfer torque on such as domain walls, skyrmions[30], and experiments have observed their motion[31,32,33]. Our purpose is to observe driving skyrmions and antiskyrmions only by the skyrmion-antiskyrmion interaction, and it is necessary to determine the range of distances from the local heating position that magnon current does not exert on the drive of skyrmions and antiskyrmions. In Fig. 3d, an electron beam with $r \sim 200$ nm was focused at a distance of $\sim 700$ nm from the antiskyrmion. After exposure in 2 minutes, although the antiskyrmion was slightly shifted away from the local heating position, magnon current is not sufficient to drive the antiskyrmion under this experimental condition (Fig. 3e), where the heat current density is inversely proportional to the distance. This experimental result indicates that magnon current does not affect skyrmion and antiskyrmions at a distance of at least more than 700 nm from the local heating position under the above experimental condition. In other words, driving skyrmions and antiskyrmions far enough away from the local heating location, more than 700 nm, should be caused only by the skyrmion-antiskyrmion interactions. Note that the antiskyrmion should have moved towards the local heating position as well as skyrmions because the skyrmion velocity model under the thermal gradient contains the square of the topological charge[30]. The magnon current response of antiskyrmions will be studied in the future. We refer to the skyrmions, antiskyrmions, and NT-bubbles collectively as magnetic particles in the following.

Applying local annihilation to the coexisting phase, we observe signatures of isotropic long-range repulsive interactions between magnetic particles. Figure 3f,g shows the particle distribution before and after the local annihilation. Figure 3f shows an image after repeated annihilation around the right region where particles are $d$ = 400-550 nm apart from each other. In turn, we exposed NT-bubble in the lower right region in Fig. 3f to the focused electron beam in 1.5 minutes. Figure 3g is the particle redistribution. The location where the electron beam was focused is labelled as FEB and the arrows indicate the direction and degree of the magnetic particle displacements from their initial positions. The ACW skyrmion closest to the annihilated NT-bubble has moved to the FEB, which cannot be ruled out by magnon current. It is noteworthy that, in the sparse region, magnetic particles at more than 860 nm from the local heating position also moved significantly, despite being unaffected by magnon current. It is clear that the driving force is the long-range interaction between magnetic particles. In Fig. 3g, magnetic particles retain a distance of 400-550 nm from each other even after displacement in the sparse region, and the particle displacements decrease with distance from the FEB. Even taking the pinning effect into account, these results can be interpreted as that the interaction is long-range repulsive (Fig. 3h). In addition, all of the magnetic particles move towards the FEB, and similar behaviour is also observed for CW skyrmions (Supplementary Fig. 6), implying that the interactions are isotropic and independent of the skyrmion helicities. Besides, Fig. 2a implies isotropic interaction between skyrmions and antiskyrmions because the distance between them almost retains at ~335 nm. However, the experiments include uncertainties due to the presence of inevitable NT-bubbles and random particle distribution.

We show the results of micromagnetic simulations supporting the experimental interpretations. Simulations used a model with only two particles, excluding inevitable NT-bubbles in the L-TEM experiments. First, we focus on skyrmion-antiskyrmion interaction. Figure 3i shows energy density as a function of $d_{Sk-aSk}$ between the ACW (CW) skyrmion and antiskyrmion at $B_{c1}/B_0$. The skyrmion and antiskyrmion are aligned axially in the [100] and [110] directions. Their energy densities decrease as the skyrmion-antiskyrmion distance increases. Figure 3j shows the corresponding conservative force (Fig. 3i). Regardless of angle and skyrmion helicities, all of the conservative forces show positive values at all distances and approach asymptotically to zero far enough away at least 500 nm (see inset). These features indicate that the force acting on each other is repulsive, long-range, and isotropic, which explain the experimental results. Our experimental and simulation results are in stark contrast to angle-and helicity-dependent short-range pairwise interactions simulated by 2D models for various magnetic systems[8,10,11,14,34,35] and other systems[1-5]. The fact that short-range pairwise interaction is not universal would be interesting in various physics and may clue the unsolved properties arising from interacting skyrmions and antiskyrmions.

We propose a model to explain the cause of the isotropic long-range repulsive interactions in terms of topological protection against the intrusion of magnetic flux density into skyrmions and antiskyrmions. We tried to visualize magnetic flux density coming from the antiskyrmion (ACW skyrmion) into the ACW skyrmion (antiskyrmion) by the transport of intensity equation (TIE) analysis[36] on the enlarged images of Fig. 3f,g with $d_{Sk-aSk}$ = ~420 nm, ~480 nm which are before and after applying local heating. However, the outside in-plane magnetic flux density ($\boldsymbol{B}_{outside}$) is too weak to be capable of visualizing it by the TIE analysis on these L-TEM images (Fig. 4a,h). The exception is in-plane $\boldsymbol{B}_{outside}$ in the near region outside the magnetization inside the antiskyrmion, which remains virtually unchanged before and after local heating. (We can visualize in-plane $\boldsymbol{B}_{outside}$ that smoothly connects antiskyrmions situated close to each other in a square antiskyrmion lattice, as shown in supplementary Fig. 7.) The simulated L-TEM images (Fig. 4b,i) and the corresponding in-plane magnetic flux density maps (Fig. 4c,j) support the experimental results. The very faint contrasts outside the skyrmion and antiskyrmion and very weak in-plane $\boldsymbol{B}_{outside}$ appear, which should be hampered by noise in the experiment. Although the measurable limit hampers visualizing $\boldsymbol{B}_{outside}$, our simulations provide insight into the role of $\boldsymbol{B}_{outside}$ in the long-range repulsive interaction. Figures 4d,k show maps of magnetic flux densities ($\boldsymbol{B}_{demag}$) corresponding to demagnetizing fields inside the thin film, and Figures 4e,l show maps of those ($\boldsymbol{B}_{leak}$) corresponding to leak magnetic fields outside the thin film. We have applied colour compensation to these maps to see weak $\boldsymbol{B}_{outside}$, as shown in Fig. 4f,g,m,n. The distribution of in-plane $\boldsymbol{B}_{demag}$ is the same as that of in-plane $\boldsymbol{B}_{leak}$, and the intensity of in-plane $\boldsymbol{B}_{demag}$ is higher than that of in-plane $\boldsymbol{B}_{leak}$. We observe that a portion of both in-plane $\boldsymbol{B}_{demag}$ and in-plane $\boldsymbol{B}_{leak}$ connects the skyrmion and antiskyrmion. A change of the relative positions of the skyrmion and antiskyrmion from $d_{Sk-aSk}$ = 420 nm to 480 nm does not significant change in distribution property of in-plane $\boldsymbol{B}_{demag}$ and in-plane $\boldsymbol{B}_{leak}$.

To clarify the role of $\boldsymbol{B}_{demag}$ and $\boldsymbol{B}_{leak}$ which connect skyrmions and antiskyrmions, we have simulated and tracked the change in the topological charge of $N$ as the ACW skyrmion and antiskyrmion approach each other beyond the closest $d_{Sk-aSk}$ = 335 nm in the experiment (Fig. 2a). Figure 5a shows the changes of $N$ as a function of $d_{Sk-aSk}$ where the ACW skyrmion and antiskyrmion are aligned axially in the [100] direction. In the range of $d_{Sk-aSk}$ > 800 nm, the skyrmion and antiskyrmion have the topological charges of $N$ = +0.993 and $N$ = –0.974, indicating the isolated skyrmion and antiskyrmion. At $d_{Sk-aSk}$ = 415 nm and 611 nm, $N$ of the antiskyrmion and skyrmion begins to deviate, respectively, (the insets of Fig. 5a), and they gradually decrease down to $d_{Sk-aSk}$ = 284 nm and 270 nm. Furthermore, below $d_{Sk-aSk}$ = 256 nm and 270 nm, $N$ of the antiskyrmion and the skyrmion change significantly, respectively. At the closest distance of $d_{Sk\text{-}aSk}$ = 241 nm, the skyrmion and antiskyrmion elongate in the [100] direction (Fig. 5b) and have topological charges of $N$ = +0.976 and $N$ = –0.973, respectively. (Below $d_{Sk\text{-}aSk}$ = 241 nm, single NT-bubbles appear, as shown in supplementary Fig. 8.) Figures 5c,d, show the in-plane $\boldsymbol{B}_{demag}$, in-plane $\boldsymbol{B}_{leak}$ maps at $d_{Sk\text{-}aSk}$ = 241 nm, respectively, where colour compensation is not applied. The in-plane $\boldsymbol{B}_{demag}$ and in-plane $\boldsymbol{B}_{leak}$ connecting the skyrmion and antiskyrmion is considerably stronger than that of Fig. 4d,e,k,l, and should strongly affect the magnetization distribution related to the topological charge of the skyrmion and antiskyrmion. Topological charge density (TCD) maps reveal the effect of the in-plane $\boldsymbol{B}_{demag}$ and in-plane $\boldsymbol{B}_{leak}$ connecting the skyrmion and antiskyrmion. Figure 5e display the TCD maps at $d_{Sk\text{-}aSk}$ = 241 nm and 818 nm, respectively. In the magnetization region where the skyrmion and antiskyrmion are closest to each other, the area of the region with finite value of TCD decreases from $d_{Sk\text{-}aSk}$ = 818 nm to 241 nm. Similar behaviours are also confirmed in the CW skyrmion and antiskyrmion aligned axially in the [100] direction and the ACW (CW) skyrmion and antiskyrmion aligned axially in the [110] direction (supplementary Fig. 9a,b). These results suggest that the isotropic long-range repulsive interaction between skyrmions and antiskyrmions arise to prevent topological destruction against the intrusion of magnetic flux density. However, although the change in the topological charge we have shown above is small and theory and experiment have shown fractional topological charges[37,38], whether continuous change of the topological charge (Fig. 5a) is reasonable from the viewpoint of topology should needed to study by theory and experiment in the future.

    We observe that the antiskyrmion-antiskyrmion interaction turns from anisotropic to isotropic. Simulations show that the skyrmion-skyrmion and antiskyrmion-antiskyrmion interactions have isotropic long-range repulsion at $B_{c1}/B_0$ as well (Fig. 6a,b). The simulation result for the skyrmions (Fig. 6a) can explain the previous experiment[21,22] and our L-TEM result showing the closest packed triangular skyrmion lattice (inset of Fig. 6a). The repulsive interaction of skyrmions has already reported in B20-type FeGe where skyrmions exist in field-induced FM state[39]. On the other hand, its previous experiment[22] has shown the square antiskyrmion lattice, indicating anisotropic interaction, inconsistent with our simulation result for the antiskyrmions (Fig. 6b). To bridge the discrepancy, we track the behaviour of the antiskyrmions with increasing $B$ (Fig. 6c). In 386 mT, the square antiskyrmion lattice is observed. (At low magnetic fields of 386 mT in Fig. 6c, antiskyrmions have rectangularly shape rather than square one because of beginning to extend in the [100] directions towards stable helical spin structures. See supplementary Fig. 3.) As $B$ is increased to 444 mT, the distribution changes to partially disordered state, with a local presence of the square lattice. At 453 mT which is around a little below the upper limit of $B_c$, the distribution is fully disordered, corresponding to the simulation result (Fig. 6b). The observed behaviours are consistent with the simulations as a function of $B$ (Supplementary Fig. 10). At 478 mT, the state turns to the isolated antiskyrmions.

In the Heusler material, we show the stochastic reversible transformation induced by a negative in-plane magnetic field and find RT coexisting phase. Our micromagnetic simulations suggest that a coexisting phase of Heusler materials can be an equilibrium state, different from theoretically-predicted non-equilibrium coexisting phases of frustrated magnets using the 2D models[10-15]. The micromagnetic simulation results suggest that the energy competition of skyrmions and antiskyrmions is sensitive to the exchange stiffness constants (chemical compositions) and the sample thickness, providing that adjusting these parameters is a way to realizing coexisting phases in this kind of Heusler materials. Previous studies reported that the Heusler materials exhibit a large topological Hall effect resulting from both momentum and real-space Berry curvatures[40] and tunable stability of the antiskyrmions by the compositions[28]. These unique features raise expectations of exhibiting versatile properties in the coexisting phases of the Heusler materials. Furthermore, our results shed light on a new-type of skyrmion-antiskyrmion interaction. Contrary to anisotropic short-range pairwise interactions between skyrmions and antiskyrmions that have been predicted in various systems[1-5,8,10,11,14,34,35], we show that skyrmions and antiskyrmions avoid each other and have an isotropic long-range repulsive interaction in the Heusler material, possibly resulting from the topological protection against the intrusion of leak magnetic flux density. Interacting skyrmions and antiskyrmions should govern the dynamic properties[39,41]. Elucidating the current-driven motion of mutually correlated skyrmions and antiskyrmions with possible quite different current responses[14,16] in the Heusler materials will be an interesting research challenge towards developing skyrmion-antiskyrmion-based spintronics.

**Methods**

A bulk polycrystalline sample of tetragonal inverse Heusler material $Mn_{1.3}Pt_{1.0}Pd_{0.1}Sn$. An alloy ingot with a nominal composition of $Mn_{1.4}Pt_{0.9}Pd_{0.1}Sn$ was prepared from high-purity Mn (99.99 wt.%), Pt (99.95 wt.%), Pd (99.95 wt.%) and Sn (99.99 wt.%) via Ar arc melting. The ingot was subsequently sealed in an evacuated silica tube and annealed at 1073 K for a week, followed by water quenching. The sample composition and thickness were determined by the energy dispersive X-ray analysis and electron energy loss spectroscopy, respectively. The thin film for the L-TEM observations was thinned by mechanical cutting and then argon ion milling method. The Fresnel mode of Lorentz transmission electron microscope (JEM2100F, JEOL) was used to observe the magnetic structures. A magnetic field was applied parallel to the incident electron beam by controlling the objective-lens current. The electron current density is about 25.0 pA/cm$^2$ when electron beam is focused.

Micromagnetic simulations were conducted using MuMax3 (Ref. 42) where we incorporate the anisotropic DMI. We performed the simulation using a micromagnetic model by the ferromagnetic exchange, uniaxial anisotropy, anisotropic DMI (Refs. 21, 22, 23), Zeeman, and demagnetizing field energies, with a size of 2560×2560×100 nm$^3$ on a 512×512×20 mesh under periodic boundary conditions for the *x-y* planes. Magnetic structures are simulated by full energy minimization using a conjugate gradient method. To calculate the demagnetizing field of the thin film, we used MuMax3 build-in function "SetPBC(10,10,0)", that is, the demagnetizing field generated by 440 copies of the simulated area was taken into account[42]. In Fig. 3i, 5, 6a,b, which were simulated using a model with only two particles, the sample size was determined so that there is no effect of periodic boundary conditions on the two particles. In Fig. 2b, 3i, 4b-g,i-n, 5, 6a,b, the material parameters are the same as in Ref. 22: the saturation magnetization $M_{sat}$ = 445 × 10$^3$ A/m, exchange stiffness constant $A$ = 8.0 × 10$^{-12}$ J/m, micromagnetic constant of the anisotropic DMI $D$ = 4.0 × 10$^{-4}$ J/m$^2$, and uniaxial anisotropy constant $K_u$ = 1.0 × 10$^5$ J/m$^3$. The normalization constant of magnetic fields is defined as $B_0 = D^2/2M_{sat}A$.

In the simulated $\boldsymbol{B}_{demag}$ and $\boldsymbol{B}_{leak}$ maps, colour vectors represent the line integral over the in-plane magnetic flux density components along the [001] direction in the thin film region of 100 nm (Fig. 4d,k and Fig. 5c) and in the vacuum region of a total of 250 nm (Fig. 4e,l and Fig. 5d) from the sample surfaces.

LTEM image simulations are based on a Fourier approach[7,43-45]. The phase of the electron beam ($\varphi$) is given by

$$\varphi(x,y) = -\frac{e}{\hbar} \int A_z(x,y,z)\,\mathrm{d}z,$$

where $\hbar$, $e$, and, $A_z$ are the reduced Planck constant, the elementary electric charge, and the *z*-component of the vector potential. $\varphi$ in reciprocal space is calculated as

$$\tilde{\varphi}(k_x,k_y) = \frac{ie}{\hbar} \frac{\widetilde{B'}_x(k_x,k_y)\,k_y - \widetilde{B'}_y(k_x,k_y)\,k_x}{k^2},$$

where $k^2 = k_x^2 + k_y^2$. $k_x$ and $k_y$ are the *x*- and *y*-components of the spatial frequency, respectively. $B'_x$

and $B'_y$ are the $x$- and $y$-components of the integration value of magnetic flux density including $B_{\text{leak}}$:

$$B'_x(x,y) = \int_{-\frac{t'}{2}}^{\frac{t'}{2}} \{\mu_0 M_{\text{sat}} \boldsymbol{m}(x,y,z) + \boldsymbol{B}_{\text{demag}}(x,y,z) + \boldsymbol{B}_{\text{leak}}(x,y,z)\} \cdot \boldsymbol{e}_x \, \mathrm{d}z$$

and

$$B'_y(x,y) = \int_{-\frac{t'}{2}}^{\frac{t'}{2}} \{\mu_0 M_{\text{sat}} \boldsymbol{m}(x,y,z) + \boldsymbol{B}_{\text{demag}}(x,y,z) + \boldsymbol{B}_{\text{leak}}(x,y,z)\} \cdot \boldsymbol{e}_y \, \mathrm{d}z.$$

$\mu_0$ is the permeability of free space. $t'$ is the integration range of magnetization and demagnetization field. The electron disturbance is calculated by

$$g(k_x, k_y) = \iint \exp\{i\varphi(x,y)\} \exp\{-2\pi i(k_x x + k_y y)\} \mathrm{d}x\mathrm{d}y.$$

By considering transfer function

$$t(k_x, k_y) = A(k_x, k_y) \exp\left[-2\pi i \left\{\frac{C_s \lambda^3 (k_x^2 + k_y^2)^2}{4} - \frac{\lambda \Delta f (k_x^2 + k_y^2)}{2}\right\}\right]$$

and Gaussian distribution

$$E_s(k) = \exp\left\{-\left(\frac{\pi \alpha}{\lambda}\right)^2 (C_s \lambda^3 k^3 + \Delta f \lambda k)^2\right\},$$

the LTEM image intensity is calculated as

$$I(x,y) = \left|\iint g(k_x, k_y) t(k_x, k_y) E_s(k) \exp\{2\pi i(k_x x + k_y y)\} \mathrm{d}k_x \mathrm{d}k_y\right|^2.$$

$A(k_x, k_y)$ is the pupil function and can be assumed to be constant for all reciprocal space. $C_s$, $\lambda$, and $\alpha$ are the spherical aberration of the objective lens, the wavelength of propagating wave, and the beam divergence angle, respectively. We used the following parameters: $\lambda$ = 2.51 pm, $t'$ = 600 nm, $C_s$ = 50 mm and $\alpha$ = 10 µrad (Refs. 46,47).

## Data availability
The data that support the results of this study are available from the corresponding authors upon reasonable request.

## Code availability
The micromagnetic simulation code (Mumax[3]) is available at https://mumax.github.io/ and the LTEM image simulation code is available from the corresponding authors upon reasonable request.

**Acknowledgements**

We thank M. Araidai, S. Harada for helpful discussions, and A. Akama, K. Higuchi, Y. Yamamoto for technical support of the TEM experiments. This work was supported by Nanotechnology Platform Project, MEXT, Japan, the JSPS KAKENHI (grant numbers 18K04679, 17H02737, and 20K20899), the JST-Mirai Program, Japan (grant number JPMJMI18G2), the joint usage/research program of the Institute of Materials and Systems for Sustainability (IMaSS), Nagoya University, and Young Researcher Grant from Center for Integrated Research of Future Electronics, IMaSS, Nagoya University.

**Author contributions**

D.S. and M.N. designed the experiment and wrote the manuscript. D.S. prepared the thin film, performed L-TEM experiments, micromagnetic simulations, EDX and EELS measurements, and analysed the experimental data. T.N. performed micromagnetic simulations. Y.G.S. synthesized a bulk polycrystalline sample. All authors discussed the data and commented on the manuscript. M.N. supervised the study.

**Competing interests**

The authors declare no competing interests.

**Corresponding author**

Correspondence to D. Shimizu and M. Nagao.

**Figure**

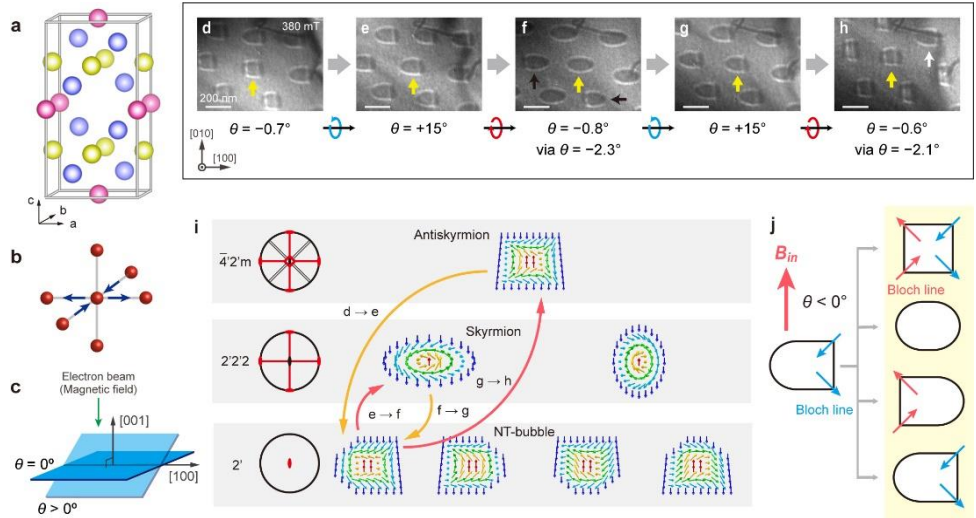

Fig. 1. **Stochastic reversible transformations between elliptical skyrmions and square-shaped antiskyrmions via NT-bubbles.** a, Crystal structure of Mn-Pt-Pd-Sn tetragonal inverse Heusler compounds depicting only the Mn atoms. The colours represent the distinguished Wyckoff positions of space group *I*-42*m* in $D_{2d}$ symmetry: magenta: 2*b*, yellow: 4*d*, blue: 8*c*. b, Schematics of anisotropic DMI. The red balls and the arrows represent the magnetic atoms and the Dzyaloshinskii–Moriya vectors, respectively. c, Sample tilt direction and tilt angle of *θ* relative to the observation direction. d-h, A series of under-focused L-TEM images of a stochastic reversible transformation between antiskyrmions and skyrmions by repeatedly tilting the sample. The tilt angles are described below the L-TEM images. The yellow arrows are distinct spin textures showing the antiskyrmion (d,h), skyrmion (f), and NT-bubble (e,g). The black and white arrows represent the changed and unchanged NT-bubbles, respectively. i, Three-dimensional magnetic point groups for antiskyrmions, skyrmions, and NT-bubbles in order of higher symmetry from the top. j, Schematics of the mechanism of a stochastic transformation from the NT-bubble (the left) to the other spin structures (the right column) by the negative in-plane magnetic fields. The schematics in the right column are the antiskyrmion, skyrmion, changed NT-bubble, and unchanged NT-bubble, from top to bottom. The large red arrow represents the direction of the negative in-plane magnetic fields ($B_{in}$) by tilting the sample to *θ* < 0° in the experiment. The small red and blue arrows represent Bloch lines of which net magnetization component is parallel and antiparallel to $B_{in}$, respectively.

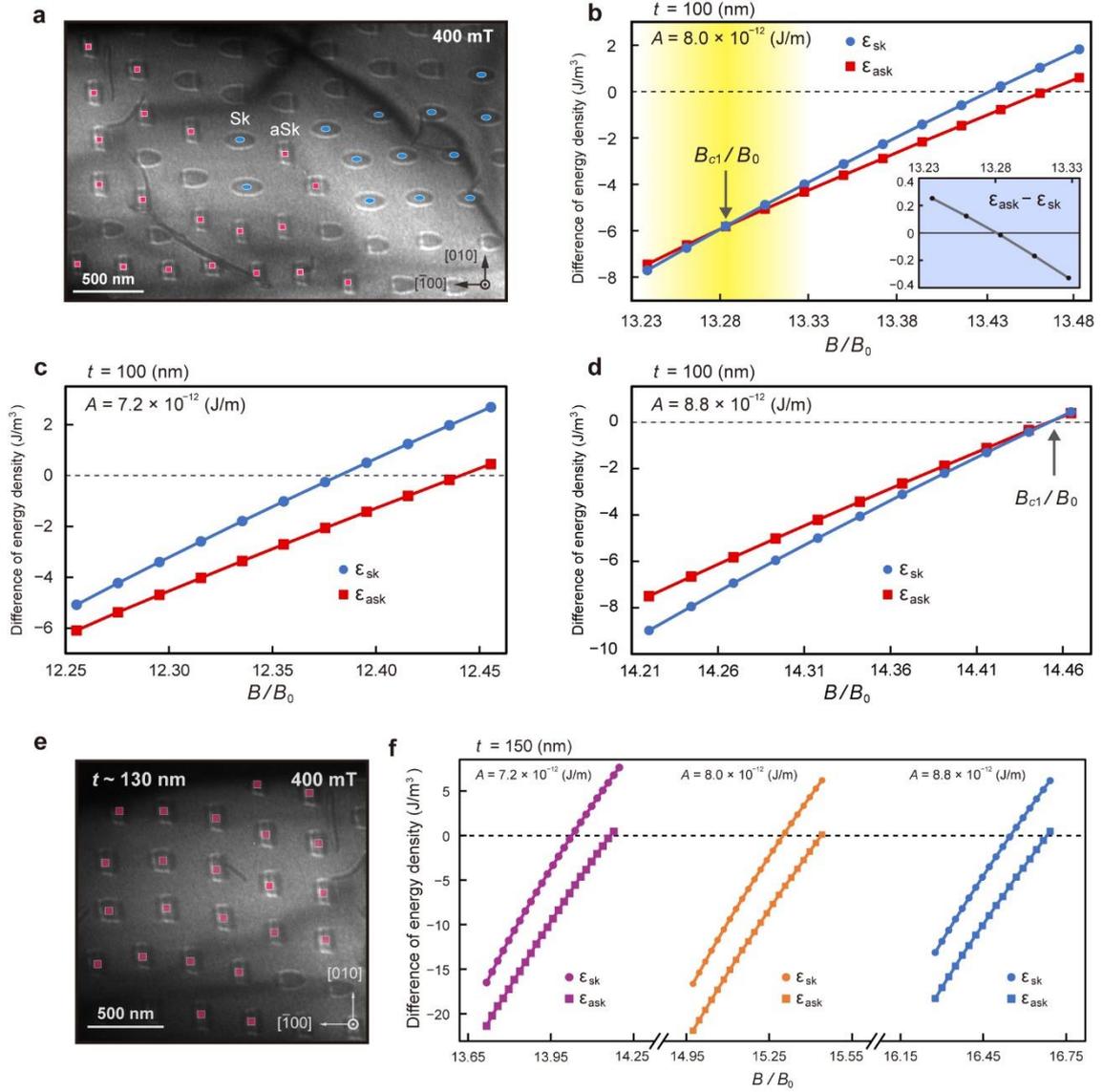

Fig. 2. Coexisting phase and energy stability of skyrmions and antiskyrmions. a, L-TEM image of the coexistence of skyrmions and antiskyrmions. Skyrmions (Sk) and antiskyrmions (aSk) are stamped by the blue ellipses and red squares, respectively. b,c,d,f, Subtracted energy density $\varepsilon_{sk}$ (blue) and $\varepsilon_{ask}$ (red) as a function of the normalized magnetic field $B/B_0$, obtained from micromagnetic simulations. $\varepsilon_{sk}$ and $\varepsilon_{ask}$ subtract the FM energy density from the skyrmion and antiskyrmion energy densities, respectively. $B_{c1}/B_0$ indicates the intersection between the $\varepsilon_{sk}$ and $\varepsilon_{ask}$ lines. $A = 8.0 \times 10^{-12}$ J/m and $t = 100$ nm (b), $A = 7.2 \times 10^{-12}$ J/m and $t = 100$ nm (c), $8.8 \times 10^{-12}$ J/m and $t = 100$ nm (d), $A = 7.2 \times 10^{-12}$ J/m (purple), $8.0 \times 10^{-12}$ J/m (orange), $8.8 \times 10^{-12}$ J/m (blue) and $t = 150$ nm (f). Inset of (b): Difference between $\varepsilon_{sk}$ and $\varepsilon_{ask}$ (i.e., $\varepsilon_{ask} - \varepsilon_{sk}$) at $B_{c1}/B_0 \pm 0.05$. e, L-TEM image obtained by a region of $t = \sim 130$ nm.

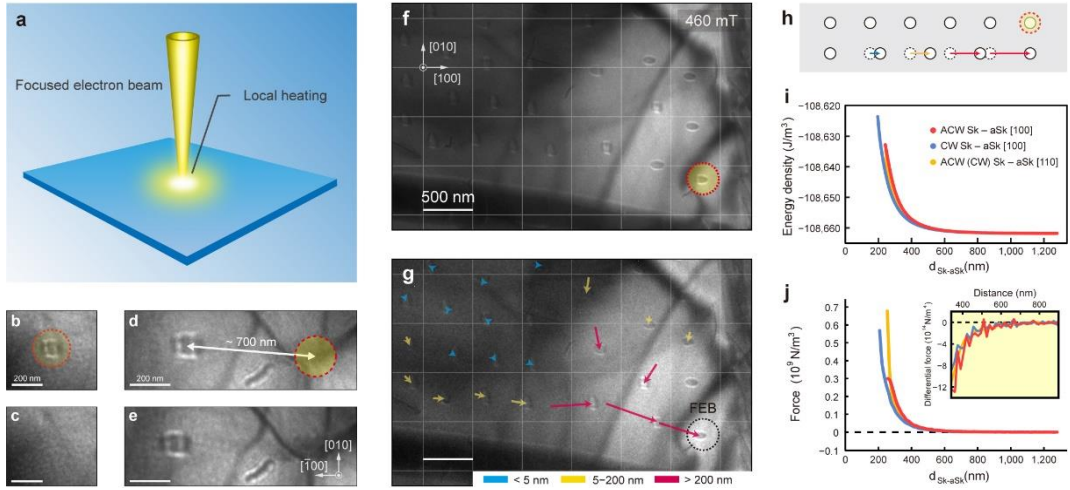

Fig. 3. Skyrmion-antiskyrmion interactions. a, Schematic of local heating with the focused electron beam. b,c, Experimental demonstration of annihilating the antiskyrmion by exposure to the focused electron beam with a diameter of $r \sim 200$ nm in 2 minutes (yellow-shaded circle). d,e, Thermal (magnon) current effect on the spin texture. Electron beam ($r \sim 200$ nm) was focused at a distance of ~700 nm from the antiskyrmion (d), and after exposure in 2 minutes (e). f,g, Particle (spin texture) distribution comparison. The cells are eye guides for the scale $500 \times 500$ nm. (f): After repeated particle annihilations in the right region. (g): Particles redistribution by annihilating NT-bubble after exposure in 1.5 minutes. FEB means the annihilation location as indicated by dotted circle. The arrows indicate the direction and degree of the particle displacements from their initial positions. h, Schematic of particle displacements in a horizontal row with repulsion. Particles are at regular intervals before (upper) and after annihilating (lower) the rightmost particle. i, Simulated energy density as a function of the ACW (CW) skyrmion-antiskyrmion distance at $B_{c1}/B_0$. The skyrmion-antiskyrmion distributions are axially aligned in the [100] direction: ACW Sk–aSk [100] (red) and CW Sk–aSk [100] (blue), and in the [110] directions: ACW (CW) Sk–aSk [110] (yellow). ACW Sk–aSk [110] is equal to CW Sk–aSk [110] (Supplementary Fig. 11). j, Corresponding conservative force calculated by the gradient of the energy density (i). Inset: Differential of the force as a function of distance.

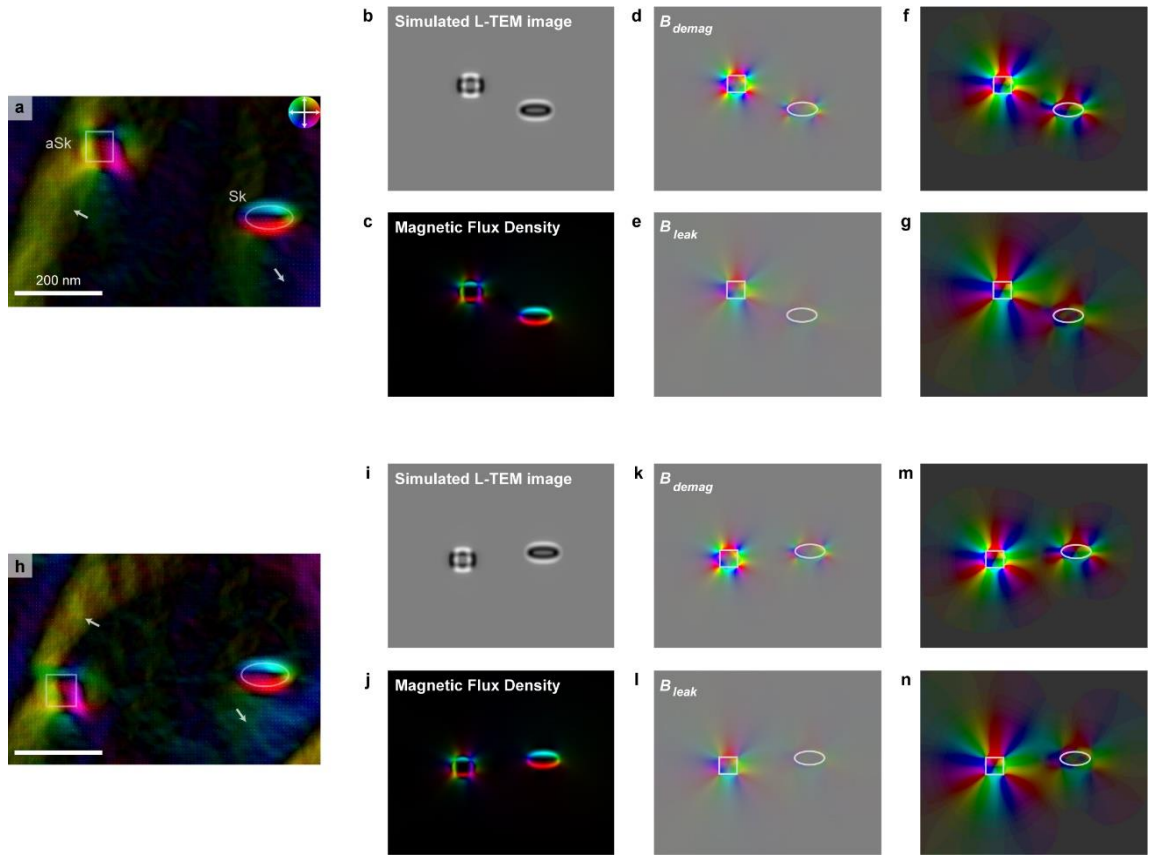

Fig. 4. Magnetic flux density maps. a,h, Magnetic flux density maps by the TIE analysis on the enlarged images of Fig. 3f,g with $d_{Sk-aSk}$ = ~420 nm, ~480 nm which are before (a) and after applying local heating (b). The greenish-yellow colours with high intensity partially covering the antiskyrmion and blue-green colours as indicated by the white arrows are false magnetic flux density originating from bend contours where the Bragg condition of the crystal is locally satisfied due to sample bend. The difference in noise between (a), and (h) is due to the difference in the electron dose during imaging before and after focusing the electron beam by local heating. b,i, Simulated L-TEM images corresponding to (a),(h). c,j, Simulated magnetic flux density maps corresponding to (b),(i). d,k, Simulated $B_{demag}$ maps inside the sample corresponding to (c),(j). e,l, Simulated $B_{leak}$ maps corresponding to (c),(j). The colour wheel (inset of (a)) represents the direction and strength of magnetic flux at every point in (a),(c),(d),(e),(h),(j),(k),(l). f,g,m,n, Maps of (f),(g),(m),(n) applied by colour compensation for (d),(e),(k),(l), respectively. Closed white lines are the outlines of the skyrmion and antiskyrmion.

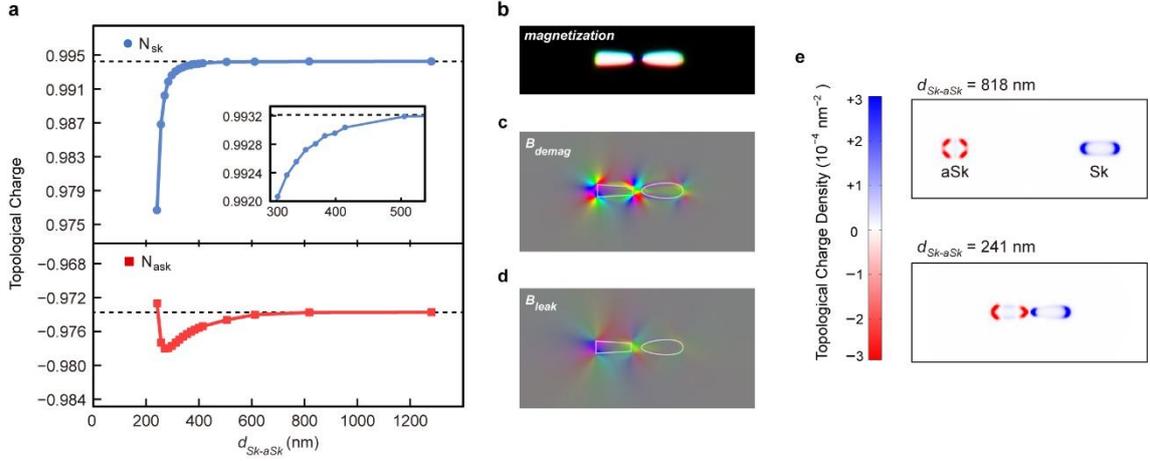

Fig. 5. Simulations of distance between the skyrmion and antiskyrmion dependence of topological charge. a, Changes of $N$ of the skyrmion (upper) and antiskyrmion (lower) as a function of $d_{Sk\text{-}aSk}$ where the ACW skyrmion and antiskyrmion are aligned axially in the [100] direction. Inset: Enlarged view ranging from $d_{Sk\text{-}aSk}$ = 300 nm to 600 nm in the case of the skyrmion. b, Simulated magnetization map at $d_{Sk\text{-}aSk}$ = 241 nm. c,d, Simulated $B_{demag}$ and $B_{leak}$ maps at $d_{Sk\text{-}aSk}$ = 241 nm. Closed white lines are the outlines of the skyrmion and antiskyrmion. e, Topological charge density map at $d_{Sk\text{-}aSk}$ = 818 nm (upper) and 241 nm (lower).

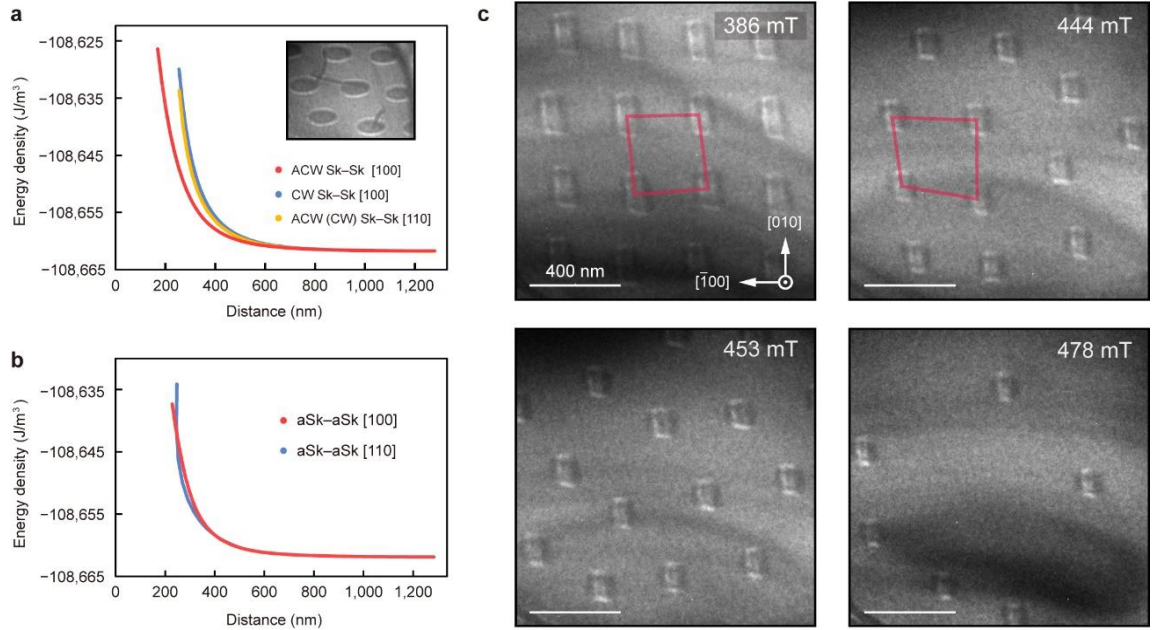

Fig. 6. Skyrmion-skyrmion and antiskyrmion-antiskyrmion interactions. a, Simulated energy density as a function of the ACW (CW) skyrmion-skyrmion distance at $B_{c1}/B_0$. The notations are similar to those in Fig. 3i. Inset: L-TEM image of a triangle lattice of elliptical ACW skyrmions. b, Simulated energy density as a function of the antiskyrmion-antiskyrmion distance at $B_{c1}/B_0$. c, A series of L-TEM images of the behaviour of the antiskyrmions with increasing $B$. Upper left: 386 mT (square lattice), upper right: 444 mT (partially disordered state with a local presence of the square lattice), lower left: 453 mT (fully disordered state), and lower right: 478 mT (isolated antiskyrmions).

**Supplemental Material**

**S1. Contrasts in the L-TEM images**

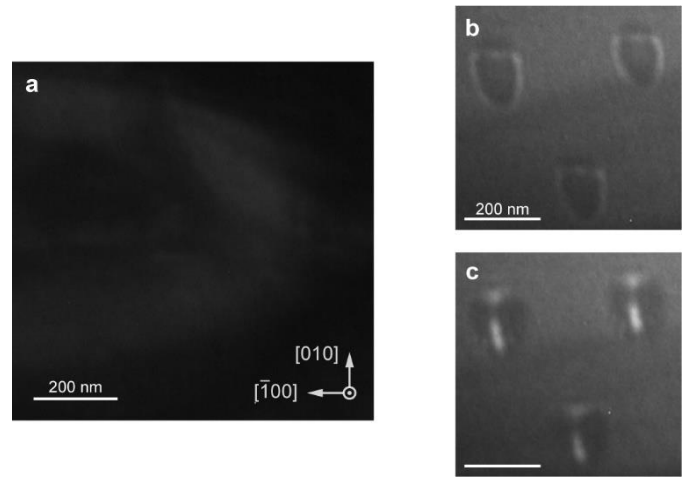

Supplementary Fig. 1. Contrasts in the L-TEM images. a, L-TEM image of the (001) thin film precisely orientating the crystal zone axis of the [001]. The image shows pitch-black due to the diffraction contrasts which obscure the magnetic contrasts. b,c, Defocused L-TEM images of the NT-bubbles. Over-focused and under-focused L-TEM images of the NT-bubbles with defocus value $\Delta f = $ +0.5 mm (b) and $\Delta f = $ 1.0 mm (c), respectively. As the defocused value is increased (i.e., the image becomes increasingly blurred), the contrast of the NT-bubbles becomes similar to that of the round antiskyrmions, which makes it difficult to distinguish between them. Therefore, it is necessary to pay attention to the defocus value in L-TEM images, and in this study, our L-TEM images were obtained with the appropriate defocus values of less than $\Delta f = \pm 0.4$ mm.

## S2. Energy stability of magnetic structures

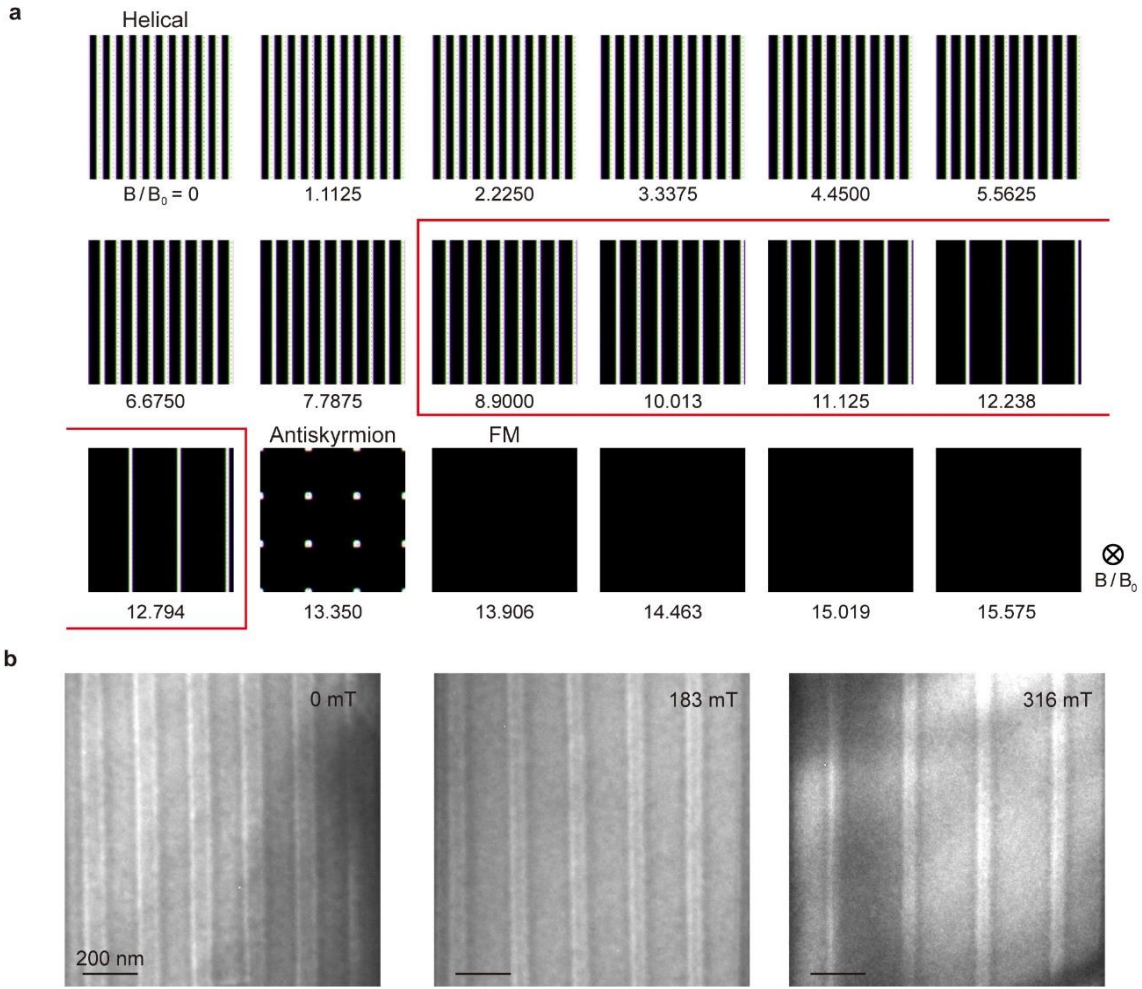

Supplementary Fig. 2. Micromagnetic simulation results of the stable magnetic structures and L-TEM images of helical spin structure depending on the magnetic fields. a, Simulated stable magnetic structures at each normalized magnetic field $B/B_0$. Each panel is the magnetic structure with lowest energy, resulting from the comparison between the periodic helical spin structure square-shaped antiskyrmion lattice, ACW(CW) elliptical skyrmion lattice, and FM. The black and white contrasts represent magnetization downwardly and upwardly perpendicular to the plane. In $B/B_0$ ranging from 8.9000 to 12.794 enclosed by the red box, the period of the helical spin structure increases with increasing $B/B_0$. b, A series of L-TEM images of helical spin structure. The period increases with increasing the magnetic field, consistent with the simulations (a, red box).

## S3. Stability of single antiskyrmions.

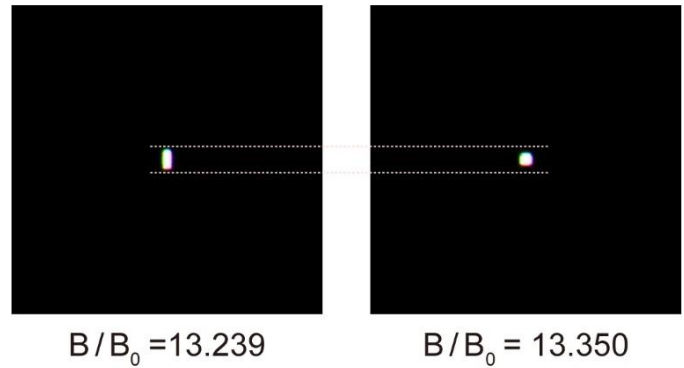

Supplementary Fig. 3. Stability of a single antiskyrmion. The simulated results of the right and left show a rectangle-shaped antiskyrmion at $B/B_0 = 13.239$ and a square-shaped antiskyrmions at $B/B_0 = 13.350$, respectively. As the normalized magnetic field decreases, the square-shaped antiskyrmion begins to extend at $B/B_0 = 13.239$ (left), indicating that the antiskyrmion state is unstable towards the helical state.

## S4. Arrangement of two magnetic particles

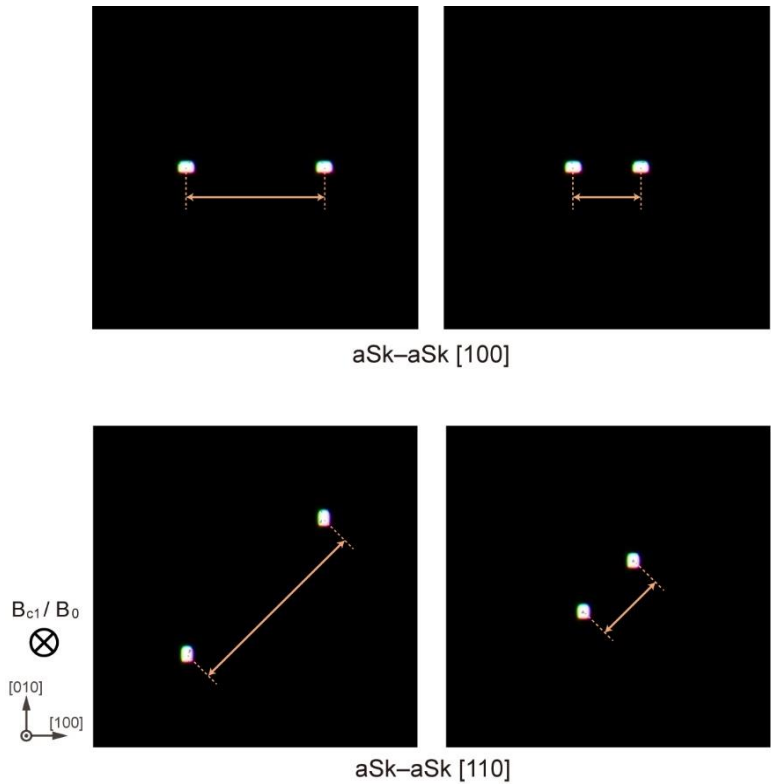

Supplementary Fig. 4. Micromagnetic simulations with model of two spin textures. Representative examples of the two antiskyrmions axially aligned in the [100] and [110] directions at $B_{c1}/B_0$. We simulated the energy density with a fixed distance between the two spin textures in the symmetric positions. Upward magnetizations were fixed for 1×1×20 cells using MuMax3 function "frozenspins", and then the spin textures are relaxed. Figures 3i, 5a and supplementary Fig. 9a show the result of the model with the antiskyrmion and ACW (CW) skyrmion. Figure 6a,b and supplementary Fig. 9b show the result of the model with the two antiskyrmions and two ACW (CW) skyrmions.

## S5. Energy stability in a thin film of $t$ = 75 nm

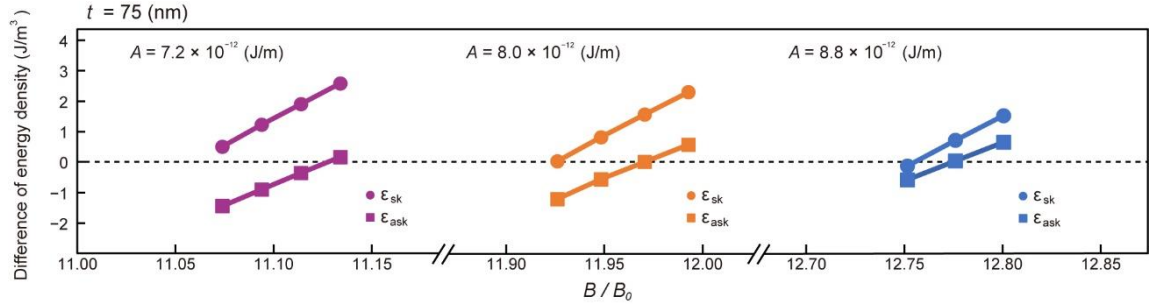

Supplementary Fig. 5. Energy stability of the skyrmion and antiskyrmion in a thin film of $t$ = 75 nm. Subtracted energy density $\varepsilon_{sk}$ (blue) and $\varepsilon_{ask}$ (red) as a function of the normalized magnetic field $B/B_0$, obtained from micromagnetic simulations. $\varepsilon_{sk}$ and $\varepsilon_{ask}$ subtract the FM energy density from the skyrmion and antiskyrmion energy densities, respectively. $A = 7.2 \times 10^{-12}$ J/m (purple), $8.0 \times 10^{-12}$ J/m (orange), $8.8 \times 10^{-12}$ J/m (blue).

## S6. CW skyrmion-antiskyrmion interactions

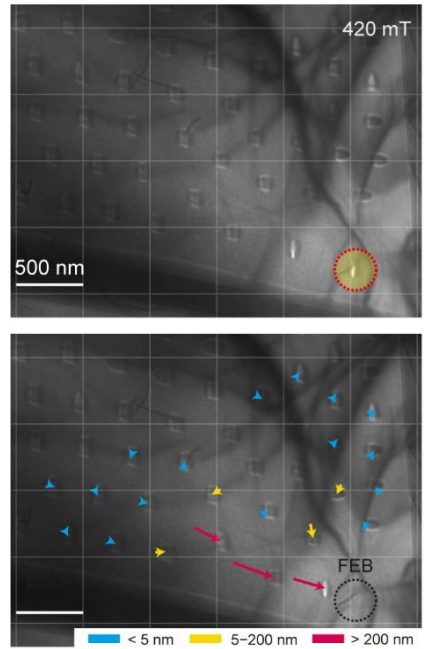

Supplementary Fig. 6. CW skyrmion-antiskyrmion interactions. L-TEM images of the distribution comparison. Upper: After repeated spin texture annihilations in the lower right region. Lower: Spin texture redistribution by annihilating the CW skyrmion after exposure to the focused electron beam in 1.5 minutes. The arrows indicate the direction and degree of the spin textures displacements from their initial positions, similar to those in Fig. 3f,g.

## S7. In-plane magnetic flux density in a square antiskyrmion lattice

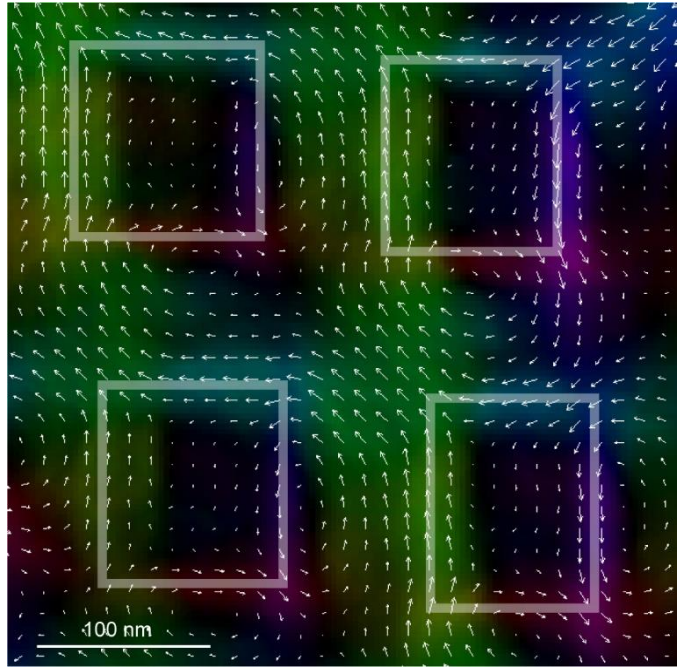

Supplementary Fig. 7. In plane magnetic flux density map of a square antiskyrmion lattice. Closed white lines are the outlines of the skyrmion and antiskyrmion. Magnetic flux density connecting antiskyrmions can be seen outside the magnetizations of antiskyrmions.

## S8. Single NT-Bubble below $d_{Sk\text{-}aSk}$ = 241 nm

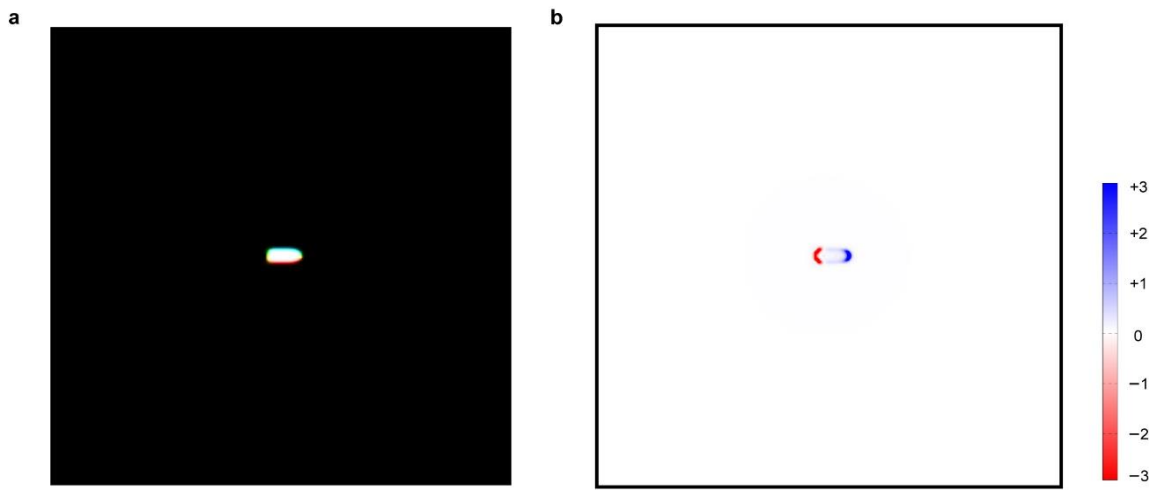

Supplementary Fig. 8. NT-bubble. a, Magnetization of a single NT-bubble. b, Topological charge density of the single NT-bubble. When the ACW skyrmion and antiskyrmion aligned axially the [100] direction below $d_{Sk\text{-}aSk}$ = 241 nm have been simulated, they are unstable and then turn into a single NT-bubble.

**S9. Simulations of distance between the skyrmion and antiskyrmion dependence of topological charge.**

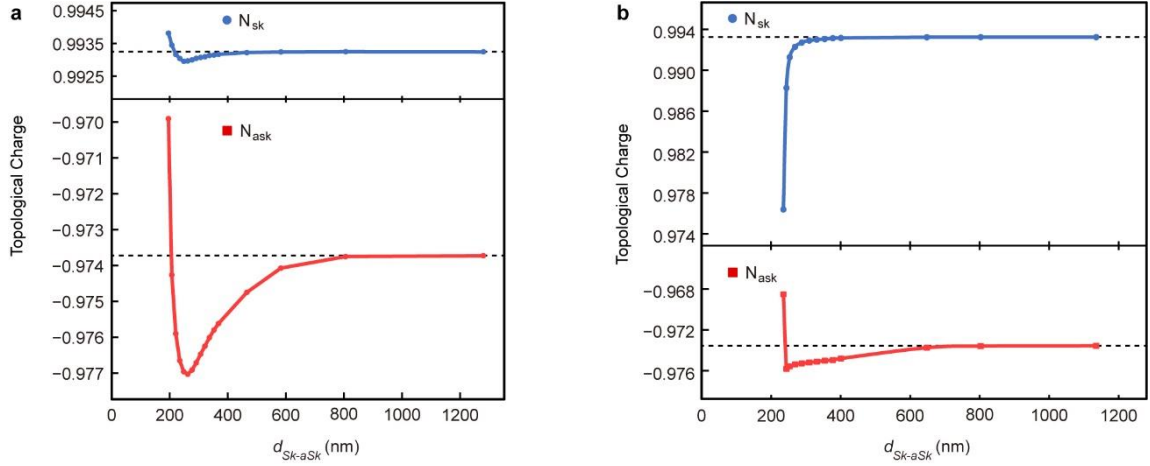

Supplementary Fig. 9. Simulations of distance between the skyrmion and antiskyrmion dependence of topological charge. a, Changes of $N$ of the CW skyrmion (upper) and antiskyrmion (lower) as a function of $d_{Sk\text{-}aSk}$ where the skyrmion and antiskyrmion are aligned axially in the [100] direction. b, a, Changes of $N$ of the CW (ACW) skyrmion (upper) and antiskyrmion (lower) as a function of $d_{Sk\text{-}aSk}$ where the skyrmion and antiskyrmion are aligned axially in the [110] direction.

## S10. Antiskyrmion-antiskyrmion interactions in micromagnetic simulation

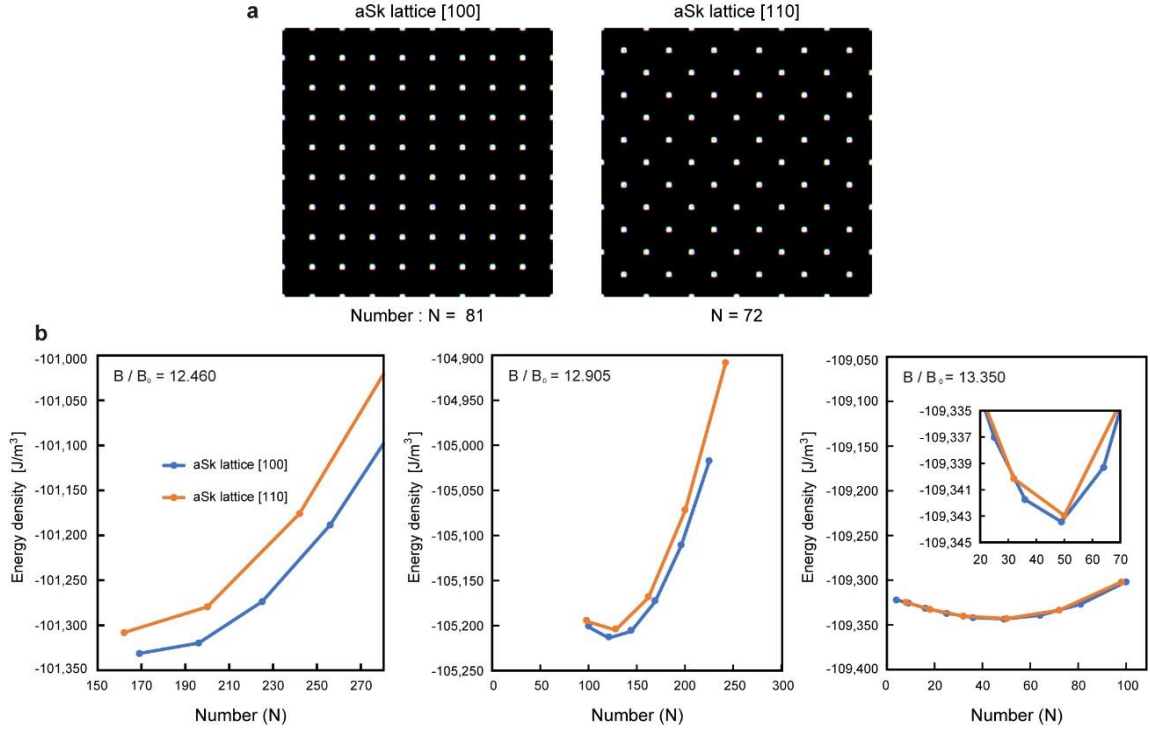

Supplementary Fig. 10. Antiskyrmion-antiskyrmion interactions in micromagnetic simulation. a, Representative examples of simulated lattice models. The "aSk lattice [100] ([110])" where the nearest antiskyrmions are aligned in the [100] ([110]) direction. The aSk lattice [100] corresponds to the square antiskyrmion lattice experimentally observed at lower magnetic field of 386 mT in Fig. 6c. The reason for the use of the lattice model is to avoid the elongation of the antiskyrmions that are metastable at lower $B/B_0$, as shown in Supplementary Fig. 3. b, Simulated energy density as a function of the number of the antiskyrmions in a size of $5120 \times 5120 \times 100$ nm$^3$ on a $1024 \times 1024 \times 20$ mesh. Left: $B/B_0$ =12.480, middle: $B/B_0$ =12.905, right: $B/B_0$ =13.350. The energy density of aSk lattice [100] is lower than that of aSk lattice [110] at $B/B_0$ =12.480 (left). As increasing $B/B_0$, the difference of the energy density between aSk lattice [100] and [110] reduces, and then is almost equal at $B/B_0$ =13.350 (inset of the right graph). Therefore, at $B/B_0$ =13.350 that is close to $B_{c1}/B_0$ = 13.283, the antiskyrmion-antiskyrmion interaction is isotropic, consistent with our L-TEM image at 453 mT in Fig. 6c.

**S11. Classifications of the arrangement of skyrmions and antiskyrmions**

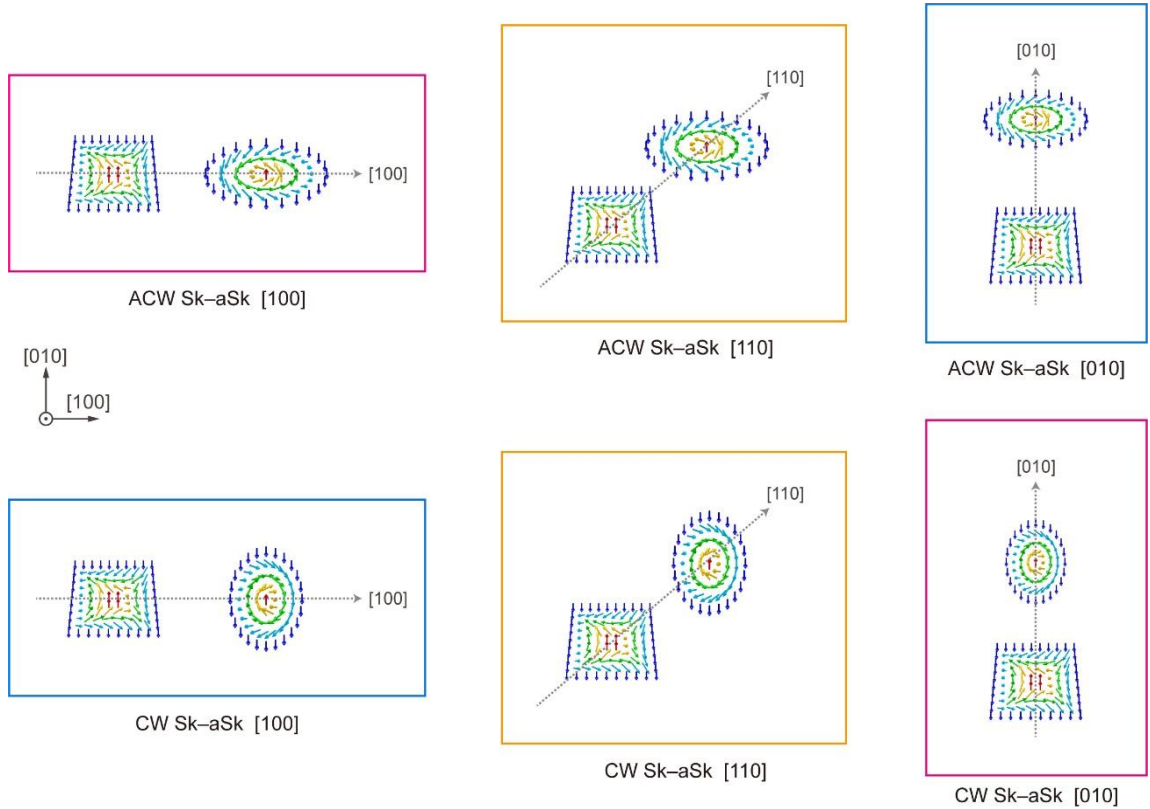

Supplementary Fig. 11. Classification of skyrmion-antiskyrmion distributions. Schematic layout of ACW (CW) elliptical skyrmion and square-shaped antiskyrmion, axially aligned in the [100], [110], and [010] directions. We simulated the skyrmion-antiskyrmion interactions to determine whether there are anisotropies or not because the spin textures have the anisotropic shapes and the skyrmion helicities. Although they appear to be different relative arrangements in the schematics, there are equivalent arrangements given their spin textures and shapes. The equivalent distributions are surrounded by the same coloured boxes. Therefore, we simulated the distributions of ACW Sk-aSk [100], CW Sk-aSk [100], and ACW Sk-aSk [110] in Fig 3i.